# Transverse geometry and physical observers

D. H. Delphenich[1]

Abstract: It is proposed that the mathematical formalism that is most appropriate for the study of spatially non-integrable cosmological models is the transverse geometry of a one-dimensional foliation (congruence) defined by a physical observer. By that means, one can discuss the geometry of space, as viewed by that observer, without the necessity of introducing a complementary sub-bundle to the observer's line bundle or a codimension-one foliation transverse to the observer's foliation. The concept of groups of transverse isometries acting on such a spacetime and the relationship of transverse geometry to spacetime threadings (1+3 decompositions) is also discussed.

MSC: 53C12, 53C50, 53C80

Keywords: transverse geometry, foliations, physical observers, 1+3 splittings of spacetime

## 1. Introduction.

One of the most profound aspects of Einstein's theory of relativity was the idea that physical phenomena were better modeled as taking place in a four-dimensional spacetime manifold $M$ than by means of parameterized curves in a spatial three-manifold. However, the correspondence principle says that any new theory must duplicate the successes of the theory that it is replacing at some level of approximation. Hence, a fundamental problem of relativity theory is how to reduce the four-dimensional spacetime picture of physical phenomena back to the three-dimensional spatial picture of Newtonian gravitation and mechanics.

The simplest approach is to assume that $M$ is a product manifold of the form $\mathbb{R} \times \Sigma$, where $\Sigma$ represents the spatial 3-manifold; indeed, most existing models of spacetime take this form. However, since this is a purely mathematical assumption, it would be more physically satisfying if it were a consequence of some more physically motivated assumption.

The most physically elementary way of accounting for one of the spacetime dimensions is to assume that $M$ is foliated by a congruence of curves, whether timelike or null, and often assumed to be geodesics; such a congruence would represent the motion of a physical observer. This approach is sometimes referred to as a *threading* of spacetime (cf., [**4**, **9**, **14**, **15**]). However, one must proceed cautiously when going about the business of showing that such a threading actually results in a product structure for $M$. In general, one must first consider the *leaf space* of the foliation, whose points each consist of a distinct curve of the foliation. Leaf spaces do not necessarily have to be differentiable manifolds, depending upon the nature of the foliation, nor do they have to define fibrations of the curves over some 3-manifold Σ or orbit spaces of the action of the proper time group $\mathbb{R}$. Hence, it would seem to be more prudent to simply deal with the leaf space on its own terms.

---

[1] E-mail: david_delphenich@yahoo.com

A complementary approach to recovering space from spacetime is to assume that M has a spacelike codimension-one foliation, whose leaves might represent proper time simultaneity submanifolds, such as when one has a well-defined proper time function $\tau: M \to \mathbb{R}$ on *M*. This approach is often referred to as a *slicing* of spacetime (cf., [**6**, **14**, **15**]). It is especially suited to the representation of Einstein's field equations for the spacetime metric tensor field *g* as a problem in time evolution, such as the ADM (Arnowitt, Deser, Misner [**1**]) formalism or the FOSH (first order symmetric hyperbolic) formalism [**18**].

Generally, either type of foliation is obtained by integration of a sub-bundle of *T*(*M*). Hence, of the two formalisms, it is generally easier to obtain a threading of spacetime, since rank-one sub-bundles of the tangent bundle *T*(*M*) to *M* – i.e., line fields – are always integrable into a global congruence of curves, which are not generally given a unique or canonical parameterization. Corank-one sub-bundles of *T*(*M*) do not have to be integrable in four-dimensional manifolds since they will be associated with a 3-form by Frobenius's theorem, and the space of 3-forms does not have to vanish in dimension four.

Hence, the focus of this study is on how one goes about describing the geometry of "space," as it is viewed by a given observer as an exercise in the transverse geometry of a one-dimensional foliation. The primary point of application to physics will be the context of spatially non-integrable cosmological models, although the methodology is sufficiently general to be applicable to essentially any physical context that involves a dynamical system.

A cosmological model, as it is often defined (cf., [**8**, **28**]), consists of a triple (*M*, *g*, **u**) in which *M* is a four-dimensional manifold that represents space-time, *g* is a Lorentzian metric tensor field of normal hyperbolic type, and **u** is a timelike unit vector field or a lightlike vector field that represents the time evolution of space, in some sense of both the terms "space" and "time."

The sense in which **u** can be defined at all generally follows from the assumption that at some sufficiently large cosmological scale the matter distribution of the Universe can be represented by a continuous time-varying spatial mass distribution. Hence, one essentially regards the evolution of the Universe at that cosmological scale as an exercise in relativistic continuum mechanics. Moreover, since the mass density of the universe at that scale is quite small at present, it is also reasonable to assume that the sort of interactions that are responsible for shear forces and viscosity in the distribution become significant only in the early stages of the Big Bang. Hence, it is also customary to regard the cosmic medium in the present epoch as a perfect fluid, which means that one can give the vector field **u** the interpretation of the relativistic flow velocity vector field of the fluid.

Any cosmological model contains the essential elements for a discussion of transverse geometry, which deals with the transverse (or normal) bundle to a foliation as the basis for all geometric constructions. However, since most of the established cosmological models involve spacetimes of the form $\mathbb{R} \times \Sigma$, for some spatial 3-manifold Σ, the question arises whether it is actually beneficial from a physics standpoint to generalize the usual geometric formalism. The answer is simply that when the foliation of *M* that is generated by the vector field **u** is spatially non-integrable, which is equivalent to saying that the vorticity vector field associated with **u** (viz., the Poincaré dual of the Frobenius 3-form $u \wedge du$, where *u* is the metric-dual covelocity 1-form) is non-vanishing the generalization is unavoidable. Although the Gödel spacetime (cf, Hawking and Ellis [**12**], which involves precisely such non-vanishing vorticity, is often regard as somewhat unphysical, due to the existence of closed timelike geodesics, which are seen as a serious breakdown in causality, nevertheless, a recurring question concerning the early history of the Universe is whether the formation of spiral galaxies suggests that at least the early history of the Big Bang might have involved a significant amount of vorticity, as well as expansion; for instance, there may have been turbulence. Hence, any such model would benefit from the methods of transverse geometry.

In section 2, we will briefly summarize the definition of a foliated manifold, give some relevant examples, and state Frobenius's theorem in the forms that we will need in the rest of the article. In section 3, we present the concept of a physical observer as essentially represented by a one-dimensional foliation; i.e., a congruence of curves. Section 4 contains a specialization of some of the elementary ideas in the transverse geometry of foliations to the case of one-dimensional foliations. In section 5, we discuss how the formalism presented here relates to the more established formalism of 1+3 splittings of spacetime; i.e., threadings. In section 6, we then summarize the key points and mention some further avenues of research.

## 2. Foliations.

A *foliation* of an *n*-dimensional differentiable manifold *M* is a partitioning $\mathcal{L}$ of *M* into a disjoint union of submanifolds of the same dimension *m* that one calls the *leaves* of the foliation. One then says that the foliation has *dimension m* or *codimension n−m*. Furthermore, one demands that all coordinate charts on *M* must be *adapted* to the foliation, in the sense that if the coordinate functions on an open subset $U \; M$ take the form $(x^i, x^a)$, with $i = 1, \ldots, m$, $a = m+1, \ldots, n$ then the intersections of *U* with the leaves of $\mathcal{L}$ can be parameterized by choosing specific values for the coordinates $x^a$. Hence, on the overlap of two such charts the coordinate transition must take leaves to leaves, and if the new coordinates are of the form $(y^j, y^b)$ then the functional form of the transformation must be:

$$y^j = y^j(x^i, x^a), \qquad y^b = y^b(x^a) . \qquad (2.1)$$

This implies that the differential map must take the form:

$$\frac{\partial y^{\mu}}{\partial x^{\nu}} = \begin{bmatrix} \dfrac{\partial y^{i}}{\partial x^{j}} & 0 \\ \dfrac{\partial y^{i}}{\partial x^{b}} & \dfrac{\partial y^{a}}{\partial x^{b}} \end{bmatrix}. \tag{2.2}$$

Hence, the differential maps take their values in the subgroup $GL(n; n-m)$ of $GL(n)$ that preserve the linear subspace defined by $(0, \ldots, 0, x^a)$.

One can define an equivalence relation on a manifold $M$ with a foliation $\mathcal{L}$ by saying that $x \sim y$ iff $x$ and $y$ belong to the same leaf of $\mathcal{L}$. The quotient space $M/\mathcal{L}$ of all such equivalence classes then represents the *leaf space* of the foliation. Since the topology of leaf spaces can be potentially pathological – in particular, they might not be differentiable manifolds – the real objective of transverse geometry is to describe the geometry of the leaf space indirectly by means of constructions that one makes on $M$ itself.

The simplest example of a foliation on a manifold is given by a product manifold $M = P \times Q$. Since either factor manifold can be regarded as representing the leaves, one can also consider the leaves of the foliation as being defined by the fibers of the relevant projection map – say, $P \times Q \to Q$. The leaf space in this case is simply $Q$.

More generally, if $p: M \to Q$ is a submersion of $M$ onto $Q$, which means that the differential map $Dp|_x$ has a rank equal to the dimension of $Q$ at every $x \in M$, then the level hypersurfaces of $p$ are submanifolds of dimension $n-q$ and collectively define a foliation of $M$ that has codimension $q$. Molino [**18**] refers to this type of foliation as a *simple* foliation, and uses it as a local model for the constructions of transverse geometry. The manifold $Q$ then serves as the leaf space of the foliation again. Of particular interest are the foliations of codimension one that one defines on a manifold $M$ by means of the level hypersurfaces of a smooth function $f: M \to \mathbb{R}$ with no critical points.

Note that foliated manifolds always behave like this example locally on the charts. Hence, every foliation is locally simple.

A more specialized case of the latter foliation is defined when the submersion $p: M \to Q$ is defined by the fibration of a manifold $M$ over a manifold $Q$ as a fiber bundle. In such a case, the leaves of the foliation are the fibers of the bundle and the leaf space is the base manifold.

In some cases, the orbits of a group action $G \times M \to M$ can define the leaves of a foliation, so the orbit space coincides with the leaf space. One calls such a foliation a *Lie foliation*. A necessary condition for a group action to foliate a manifold is that the orbits all have the same dimension, which suggests that the isotropy subgroups of all the orbits must have the same dimension, as well. For instance, the action of $O(n)$ on $\mathbb{R}^n - 0$ foliates it by $n-1$-spheres, which all have isotropy subgroups conjugate to $O(n-1)$. An *almost-free* group action, for which the isotropy subgroups are all discrete, will also define a foliation by its orbits.

The apparent origin of the sequence of mathematical generalizations that led to the definition of foliated manifolds is in the foliation of spaces by the integral submanifolds of differential systems that are defined on these spaces. In the context of manifolds, a *differential system* of rank *m* on a manifold *M* is a sub-bundle $\mathcal{D}$ of the tangent bundle that has constant rank *m* everywhere. Hence, the fiber $\mathcal{D}_x$ at $x \in M$ represents an *m*-dimensional linear subspace of $T_x(M)$ that one calls an *integral element*. A submanifold $S \to M$ of dimension $k \leq m$ is called an *integral submanifold* of $\mathcal{D}$ iff the tangent space $S_x$ to $S$ at each $x \in S$ is contained in $\mathcal{D}_x$. $\mathcal{D}$ is called *integrable* iff every $x \in M$ has some integral submanifold through it, and, in particular, *completely integrable* if all the integral submanifolds have maximum dimension equal to the rank of $\mathcal{D}$. Hence, a differential system $\mathcal{D}$ on $M$ is completely integrable iff it is the tangent bundle to a foliation of $M$ by leaves whose dimension equals the dimension of the fibers of $\mathcal{D}$.

The simplest differential systems are of rank one, and are thus defined by line fields. As we shall discuss in the next section, a line field *L* on a manifold *M* does not have to be generated by all scalar multiples of a global non-zero vector field **u** on *M* – indeed, such a vector field might not even exist – but when it does, the integral submanifolds, which are then integral curves $\gamma: \mathbb{R} \to M$, are the solutions to the first-order system of *n* ordinary differential equations:

$$\left. \frac{d\gamma(\tau)}{d\tau} \right|_{\tau=\tau_0} = \mathbf{u}(\gamma(\tau_0)). \tag{2.2}$$

In a local coordinate chart $(U, x^\mu)$, $\mu = 1, \ldots, n$ they then take the form:

$$\frac{dx^\mu}{d\tau} = u^\mu(x^\nu(t)), \tag{2.3}$$

in which the components $u^\mu$ of **u** are taken in the natural frame field $\partial_\mu = \partial/\partial x^\mu$ on *U* that is defined by the partial derivatives with respect to the coordinate functions.

Often, the differential system $\mathcal{D}$ is itself the algebraic solution to a system of exterior differential equations of the form:

$$\theta^\alpha = 0, \qquad \alpha = 1, \ldots, m, \tag{2.4}$$

in which each $\theta^\alpha$ is an exterior differential *k*-form for some *k*. One calls such differential systems *exterior differential systems*.

To say that $\mathcal{D}$ is a solution to such a system is to say that when each of the exterior forms $\theta^\alpha$ in the system is evaluated on vector fields that take their values in the fibers of $\mathcal{D}$ the result is zero. One then says that the fibers of $\mathcal{D}$ are the *annihilating subspaces* of all the forms in the system. When all of the exterior forms in the system are 1-forms, one calls the system a *Pfaffian system*. In particular, when one has a non-zero 1-form $\theta$ on a manifold *M* the annihilating subspaces are the integral elements of a differential system of corank one and the exterior differential system is simply:

$$\theta = 0. \tag{2.5}$$

The necessary and sufficient conditions for the complete integrability of a differential system $\mathcal{D}$ on a manifold $M$ are given by *Frobenius's theorem,* which says that $\mathcal{D}$ is completely integrable iff it is *involutive*. By this, we mean that the vector space $\mathfrak{X}(\mathcal{D})$ of sections of the sub-bundle $\mathcal{D} \to M$, which are then vector fields on $M$ with values in the fibers of $\mathcal{D}$, must also be a Lie sub-algebra of the Lie algebra $\mathfrak{X}(M)$ of all vector fields on $M$. That is, if $\mathbf{v}, \mathbf{w} \in \mathfrak{X}(\mathcal{D})$ then $[\mathbf{v}, \mathbf{w}] \in \mathfrak{X}(\mathcal{D})$.

This has the immediate consequence that any rank one differential system must be completely integrable, since one can then represent $\mathbf{v} = f\mathbf{u}$, $\mathbf{w} = g\mathbf{u}$ (at least locally), which then makes:

$$[\mathbf{v}, \mathbf{w}] = (f\mathbf{u}g - g\mathbf{u}f)\, \mathbf{u}. \tag{2.6}$$

Note, in particular that this bracket does not have to vanish, even though the fibers are one-dimensional, since the Lie algebra $\mathfrak{X}(\mathcal{D})$ is defined over the *infinite*-dimensional vector space of smooth functions on $M$, such as $f$ and $g$. Indeed, we could just as well write (2.6) in the form:

$$[f, g] = f\mathbf{u}g - g\mathbf{u}f. \tag{2.7}$$

Hence, any vector field on $M$ defines a Lie algebra on $C^\infty(M)$.

When a differential system is defined by an exterior differential system of the form (2.4) the form that Frobenius's theorem takes is to say that the exterior differential system (2.4) is completely integrable iff either:

$$\theta^\alpha \wedge d\theta^\alpha = 0, \qquad \text{for all } \alpha \tag{2.8}$$

or there are 1-forms $\eta^\alpha_\beta$ for each $\alpha$ such that:

$$d\theta^\alpha = \eta^\alpha_\beta \wedge \theta^\beta \qquad \text{for all } \alpha. \tag{2.9}$$

In particular, any $k$-form $\theta$ defines a completely integrable differential system by way of $\theta = 0$ iff:

$$\theta \wedge d\theta = 0. \tag{2.10}$$

## 3. Physical observers.

A *physical observer* is defined by a congruence $\mathcal{O}$ of smooth curves in the space-time manifold $M$ that represent a physical motion. That is, one has a one-dimensional foliation of some region in *M*. The nature of this region can be that of a world-tube, as in the case of a spatially extended, but bounded, distribution of mass – or at least energy, in the case of photons – or perhaps all of *M*, as in the case of cosmological models. Although it would also be prudent to leave open the possibility that there are degenerate curves – i.e., fixed points – in this congruence, that would expand the scope of the preset discussion to that of singular foliations, which we shall treat in a later investigation. For the sake of simplicity, we confine our attention in the sequel to only that region of *M* in which the congruence exists, or − what amounts to the same thing − we assume that *M* is completely foliated by $\mathcal{O}$.

By differentiation, each curve of the congruence is associated with a non-zero velocity vector field **u**($\tau$), where $\tau \in \mathbb{R}$ represents the curve parameter. If, as is usually the case in the general relativistic treatment of space-time, *M* is endowed with a Lorentzian structure, which we describe by a second-rank doubly-covariant tensor field *g* that is symmetric, non-degenerate, and globally of hyperbolic normal type (with signature type (+ − − − )) then the world lines of massive matter are characterized by $\tau$ being the proper time parameterization, viz., the one for which $g(\mathbf{u}, \mathbf{u}) = 1$, and one then calls the congruence *timelike*.

If the congruence consists of *lightlike* curves, for which $g(\mathbf{u}, \mathbf{u}) = 0$ then the proper time parameterization is impossible. For such a *null* congruence, which might describe the photons of a laser beam or the radiation from a star, one generally chooses an *affine parameterization*, which are then ones for which the geodesic equation takes the form $\nabla_\mathbf{u} \mathbf{u} = 0$. These parameterizations are not unique, but represent an equivalence class under action of the one-dimensional affine group on $\mathbb{R}$, namely: $A(1) \times \mathbb{R} \to \mathbb{R}$, ($[a, b]$, $x) \mapsto ax + b$.

In the general case, one associates each curve of $\mathcal{O}$ with the tangent line $L_x \in T_x(M)$ at each of its points, which then defines a rank-one vector sub-bundle *L* of the tangent bundle *T*(*M*); i.e., a *line field* on *M*. Although this generalization is not as necessary in the case the motion of pointlike matter, which involves only one smooth curve in *M*, in the case of extended matter the issue of whether one can actually give all of the curves of the congruence a common parameterization is more physically and mathematically involved that it might seem at first glance. We shall elaborate upon this shortly.

More precisely, since one moral principle for physics research is to experiment locally and theorize globally we shall start with the locally defined construction that *L* represents as being the fundamental object.

From Frobenius's theorem, the line bundle *L* is always integrable into a one-dimensional foliation of integral curves. However, this does not imply a number of stronger statements that sometimes get assumed in the process:
  i. The existence of a global section to *L*.
  ii. The existence of a global flow for such a global section of *L*, if it exists.
  iii. The existence of a global slice to $\mathcal{O}$.
  iv. The existence of a unique global complementary sub-bundle $\Sigma$ to *L*.
  v. The existence of a global transverse foliation to $\mathcal{O}$.

We shall now clarify the precise meaning of these comments.

The first issue in the list above amounts to the question of time-orientability. The congruence $\mathcal{O}$ is said to be *time-orientable* iff the line bundle $L$ is orientable. This, in turn, is equivalent to the existence of a global non-zero vector field **u** on $M$ that takes its values in the fibers of $L$; i.e., a global non-zero section **u**: $M \to L$ of the vector bundle $L \to M$; we shall denote the Lie algebra of all such sections by $\mathfrak{X}(L)$. If $M$ is compact then a global non-zero vector field of any description can exist iff the Euler-Poincaré characteristic $\chi[M]$ vanishes. However, this necessary condition for the existence of a non-zero section of $L$ is not sufficient.

A further necessary, but not sufficient, condition for time-orientability in the compact case is that the first Stiefel-Whitney class $w_1[L] \in H^1(M; \mathbb{Z}_2)$ of the vector bundle $L$ – which is not to be confused with the first Stiefel-Whitney class of $T(M)$ – must vanish. One notes that if $M$ is simply connected then $H^1(M; \mathbb{Z}_2)$ will vanish regardless of the choice of $L$ and any line field on a simply connected $M$ will be time orientable. Since any non-simply connected $M$ has a simply connected covering manifold $\bar{M}$ one sees that any non-time orientable congruence $\mathcal{O}$ will have a time orientable covering congruence in $\bar{M}$.

Now suppose that $\mathcal{O}$ is time-orientable and $L$ is generated by a non-zero velocity vector field **u**. In general, the assumption that **u** is continuously differentiable will only imply the existence of local flows. That is, about any $x \in M$ one will have an action of some subset $(-\varepsilon, +\varepsilon) \in \mathbb{R}$ of the group $(\mathbb{R}, +)$ on some open neighborhood $U$ of $x$:

$$\Phi: (-\varepsilon, +\varepsilon) \times U \to M, \qquad (\tau, y) \mapsto \Phi_\tau(y),$$

such that $\Phi_\tau: U \to M$ is a diffeomorphism onto its image for every $\tau \in (-\varepsilon, +\varepsilon)$. By definition, this action has the property that the orbit of any $y \in U$ will be an integral curve of the vector field **u**:

$$\mathbf{u}(y) = \left.\frac{d\Phi_\tau(y)}{d\tau}\right|_{\tau=0}.$$

One must be careful when applying the basic group property of the action:

$$\Phi_\tau \Phi_\sigma = \Phi_{\tau+\sigma},$$

since this property only applies to proper time invariant vector fields, which corresponds to the case of steady flow in hydrodynamics and autonomous systems of ordinary differential equations. However, in the case of a proper time varying vector field, **u**: $\mathbb{R} \times M \to T(M)$, one can extend **u** to a vector field $\tilde{\mathbf{u}}$ on $\mathbb{R} \times M$ in such a way that the extended system is autonomous by defining $\tilde{\mathbf{u}}(\tau, x) = (1, \mathbf{u}(t, x))$. This is essentially the Galilean embedding of velocity vectors that one encounters in the non-relativistic limit of special relativity.

In general, local flows do not extend to flows for which $U = M$, nor can one extend from the subset $(-\varepsilon, +\varepsilon)$ to all of $\mathbb{R}$; if both extensions are possible, one calls the flow *global*. A global flow will always exist in the case of a compact $M$, as well as the case of a linear **u**, which implies that $M = \mathbb{R}^n$. This shows that compactness is a sufficient, but not necessary condition for the existence of a global flow.

Issue *iii* says that one cannot generally choose a unique point $\chi(0)$ from each curve $\chi(\tau)$ in $\mathcal{O}$ that might serve as the proper time origin. Such an association defines a *global slice* of the foliation, that is, a codimension-one submanifold $S$ of $M$ that intersects $\mathcal{O}$ transversely; hence, $T_x(S) \oplus L_x = T_x(M)$ at each $x \in M$. Because this would define the set of all points in space-time that are proper time simultaneous at $\tau = 0$ for $\mathcal{O}$, one can see that such a construction must have a distinctly non-relativistic character to it. Hence, a different foliation would generally have a different global slice, if it had one at all.

Furthermore, even if **u** has a global flow there is nothing to say that there is a global slice to it. One has to note that any point $x \in M$ can serve as the proper time origin along the integral curve through it as well as any other. Hence, although $(-\varepsilon, +\varepsilon)$ has a distinguished point in the form of 0, $M$ does not, and neither do any of the integral curves of $\mathcal{O}$. Again, the issue of synchronizing the proper time parameters for all of the curves in $\mathcal{O}$, even if they are all given the unit-speed parameterization for a given Lorentzian structure, is not something that even arises in the case of the motion of pointlike matter, which is what occupies a lot of the discussion in general relativity at its most elementary level.

A stronger requirement than the existence of a single global slice at $\tau = 0$ is the existence of a global slice for each $\tau$. This would define a codimension-one foliation of $M$ transverse to the observer $\mathcal{O}$. Whether or not such a transverse foliation exists depends entirely upon the nature of the observer. If such a transverse foliation exists then one has a corank-one vector sub-bundle $\Sigma$ of $T(M)$ that is complementary to $L$ and is defined by the tangent spaces to the leaves of the transverse foliation. Each fiber $\Sigma_x$ of $\Sigma$ is defined by an equivalence class $[\alpha]$ of 1-forms $\alpha$ that annihilate all of the vectors in $\Sigma_x$, so $\alpha(\mathbf{v}) = 0$ for all $\mathbf{v} \in \Sigma_x$. Hence, the vector bundle $\Sigma$ can also be represented by a line field $\Sigma^*$ in $T^*(M)$.

One easily sees that if $\mathcal{O}$ admits a global slice $S$ and a global flow then it will admit a transverse foliation. In fact, one will then be able to represent $M$ as $\mathbb{R} \times S$. This is the usual consequence of the initial-value formulation of gravitation, since one starts with a global slice in the form of a maximal Cauchy hypersurface and hopes that its time evolution by a one-parameter group of diffeomorphisms will generate the rest of $M$.

As long as one deals with locally defined constructs, as is commonly the case in most of general relativity, one essentially has such a representation of $M$, or really, the open subset $U$ on which a coordinate chart has been defined, as $\mathbb{R} \times S$. For such a foliation, all of the issues above represent natural properties of such an elementary type of foliation. However, in this study of the geometry of physical observers, we shall not assume the existence of transverse foliations or global slices, but deal with the congruence of curves as the fundamental object.

As a consequence of this assumption, we shall need to examine some of the other special cases of a one-dimensional foliation $\mathcal{O}$ of *M*. A recurring issue is whether one can use the quotient space $M/\mathcal{O}$ of *M* by the relation $x \sim y$ iff there is an integral curve of $\mathcal{O}$ that passes through both *x* and *y*. This is, by definition, the *leaf space* of $\mathcal{O}$. In general, it not even a manifold, but sometimes one sees it represented as either a fibration or an orbit space, so we need to examine these possibilities more closely.

There are two possible ways of representing $M/\mathcal{O}$ as a fibration depending upon what one regards as the base manifold of that fibration, proper time or "space." One sees two immediate problems with this construction:

First, in either case, whether one regards the curves of $M/\mathcal{O}$ as fibers or a base manifold, they must all be diffeomorphic. However, even in the absence of fixed points, the curves might be either open − hence diffeomorphic to $\mathbb{R}$ − or closed − hence, diffeomorphic to the circle $S^1$. For any congruence in a manifold *M* there is always a covering manifold $\tilde{M}$ with a covering congruence whose integral curves are all open (see, Fried [**11**]). Although the non-existence of closed timelike curves is a common assumption of most cosmological models, they do appear in some widely studied models, such as the Gödel model. The possibility of all of the curves of the congruence being closed seems less physically motivated than all of them being open, but this might be unavoidable when considering the conformal compactification of *M*.

Second, suppose that one wishes to fiber *M* over $M/\mathcal{O}$ as a three-dimensional manifold. Hence, all of curves in $\mathcal{O}$ must have the same diffeomorphism type in order to make them the fibers of $M \to M/\mathcal{O}$. Even if we are assuming that $M/\mathcal{O}$ is a manifold that still would not necessarily imply the existence of a global slice to the congruence, since that would amount to a global section of the fibration. However, if we are assuming that all of the curves of $\mathcal{O}$ are open − hence, contractible spaces – such a global section will exist from basic obstruction theory (see Steenrod [**23**] or Milnor and Stasheff [**17**]). Such a section is by no means canonical, which can be traced to the fact that the fibers are assumed to be non-canonically diffeomorphic to $\mathbb{R}$. In the event that one assumes that all of the fibers are (non-canonically) diffeomorphic to $S^1$, whose only non-trivial homotopy group is $\pi_1(S^1) = \mathbb{Z}$, the existence of a global section will be obstructed by the possible non-vanishing of a $\mathbb{Z}$-cocycle in dimension two.

One must realize that even though it is customary in relativity theory to regard closed timelike curves in space-time as unphysical, one must keep in mind two possible ways that they might be simply *unobservable*: First, the period of the orbit could be of cosmological magnitude – probably much larger than the present age of the universe. Second, if one thinks in terms of extended matter instead of pointlike matter, there is nothing to say that the period is the same for all of the orbits of $\mathcal{O}$, or even locally the same. Conceivably, by the time a chosen point *P* on one orbit returned to that point, the points that were in a neighborhood of *P* at the first time point might not be close to their return points on a local slice through *P* when *P* returns. Hence, the chosen reference point might not actually observe a duplication of a previous state of all the neighboring matter, except for some even larger "synodic" period. One can consider the orbits of the planets as a motivating example.

Furthermore, the existence of closed timelike curves is not actually inconsistent with time orientability, even though it does imply an acausal nature to events along a closed curve. A breakdown of time orientability would have to involve a curve through a reference point that goes into the future with a future-pointing velocity vector and comes back to that point with a past-pointing velocity vector, not merely returning to it by means of curve that comes from the past at that point. For instance, one would expect that there would be a breakdown of time orientability if **u** were allowed to have zeroes. However, this might take the form of having all of the directions at such a point pointing into the future – e.g., if the zero were a source – or into the past, as in the case of a sink. This is essentially what happens at the north pole of the Earth, where all directions point south.

Clearly, $\mathcal{O}$ cannot be an orbit foliation unless one has a complete global flow for **u** and **u** has no fixed points, which means **u** has no zeroes. Furthermore, an orbit must always have a discrete isotropy subgroup in this case. In the case of an open orbit, the isotropy subgroup is 0, which makes the orbit diffeomorphic to $\mathbb{R}$, and the action is referred to as *simply transitive*. In the closed case, the isotropy subgroup is $\mathbb{Z}$, which makes the orbit diffeomorphic to $\mathbb{R}/\mathbb{Z}$, which can be regarded as either the one-torus $T^1$ or the circle $S^1$. In order for a global action of $\mathbb{R}$ to not produce a foliation, one would have to either allow fixed points or possibly something more exotic, such as a non-discrete isotropy subgroup – say, the rational numbers $\mathbb{Q}$. Of the two possibilities, considering vector fields with zeroes seems more physically motivated; however, we shall not pursue that option in the present study.

## 4. The transverse geometry of a one-dimensional foliation.

The basic approach of transverse geometry is to define geometric objects on a manifold *M* that has been given a foliation $\mathcal{F}$ in such a manner that they will behave like objects that could be defined on the leaf space $M/\mathcal{F}$ if it had been given the structure of a differentiable manifold. Such objects will then have to be "projectable" under the quotient map $M \to M/\mathcal{F}$, in a sense. Hence, it will generally be necessary for them to be, in some sense, "constant" on the leaves. Since we are only concerned with a one-dimensional foliation $\mathcal{O}$ in the present discussion, we shall specialize the more general definitions that are given for *p*-dimensional foliations (cf., Molino [**18**]) to that case.

The result of these definitions and constructions will be to allow us to speak of "spatial geometry" without the necessity of having to require that the foliation $\mathcal{O}$ actually define a product structure $\mathbb{R} \times \Sigma$ for $M$, or even a submersion $M \to \Sigma$, more generally. When one is dealing with non-integrable cosmological models, for which the vorticity vector (i.e., the Frobenius 3-form) of $\mathcal{O}$ is non-vanishing, this is essential.

It is simple enough to define constancy on the leaves for smooth functions, of course. A smooth function $f \in C^\infty(M)$ is called *basic* if any of the following equivalent conditions are satisfied:

  i.  It is constant on the leaves of $\mathcal{O}$.
  ii. $\mathbf{u}f = 0$ for any $\mathbf{u} \in \mathfrak{X}(\mathcal{O})$.
  iii. In any adapted local coordinate chart $(U, x^\mu)$ one has: $f = f(x^i)$, $i = 1, 2, 3$.

If the foliation $\mathcal{O}$ were simple and defined by a submersion $p: M \to \Sigma$ then basic functions on $M$ would all represent the pullbacks of smooth functions on $\Sigma$ by the submersion $p$; i.e., $f = \bar{f} \cdot p$ for some unique $\bar{f} \in C^\infty(\Sigma)$. We shall denote the ring of all basic functions on $M$ by $C_b^\infty(M)$.

In the absence of a connection on $T(M)$, one must do more work to define the concept of constancy on the leaves for a vector field on $M$. However, we can always use the Lie derivative, which means that we are essentially using convection along a flow to substitute for parallel translation along a curve. However, it not desirable to require the vanishing of that Lie derivative, since we wish to avoid committing ourselves to a specific choice of generating vector field $\mathbf{u}$ for $L$, or even its orientability. Since a change of parameterization for any integral curve $\gamma$ would replace $\mathbf{u}$ with $\lambda\mathbf{u}$, where $\lambda$ is a non-zero function along $\gamma$, one sees that if $\mathbf{X} \in \mathfrak{X}(M)$ and $[\mathbf{X}, \mathbf{u}] = 0$ then one would have:

$$[\mathbf{X}, \lambda\mathbf{u}] = (\mathbf{X}\lambda)\mathbf{u} . \tag{4.1}$$

Hence, we say that the vector field $\mathbf{X}$ is *foliate* (or *projectable*) iff:

$$L_\mathbf{u}\mathbf{X} = \alpha\mathbf{u}, \tag{4.2}$$

for any $\mathbf{u} \in \mathfrak{X}(\mathcal{O})$ and some $\alpha \in C^\infty(M)$ that generally depends upon $\mathbf{u}$.

Note that if $\lambda$ is any basic function on $M$ and $\mathbf{X}$ is a foliate vector field then for any $\mathbf{u} \in \mathfrak{X}(\mathcal{O})$:

$$[\lambda\mathbf{X}, \mathbf{u}] = \lambda\alpha\mathbf{u} \tag{4.3}$$

for some $\alpha \in C^\infty(M)$. Hence, the set $\mathfrak{X}_\mathcal{O}(M)$ of all foliate vector fields on $M$ is a $C_b^\infty(M)$–module. Furthermore, foliate vector field act on basic functions as derivations, just as for the vector fields on $M$ do. However, unlike the latter more general case, the representation of a derivation on $C_b^\infty(M)$ by a foliate vector field is not unique. In particular, any $\mathbf{X} \in \mathfrak{X}(\mathcal{O})$ will give the zero derivation.

An equivalent condition for $\mathbf{X}$ to be foliate is that in any adapted local coordinate chart $(U, x^\mu)$ the component functions $X^\mu$ with respect to the natural frame field must take the form $X^\mu(x^i)$, $i = 1, 2, 3$. Hence, the condition that we gave satisfactorily generalizes the notion that the vector field should be transverse to the foliation.

If a covector field $\alpha$ on $M$ – i.e., a 1-form – annihilates any vector tangent to the foliation, then it is tempting to use this as a definition of its transversality. However, since one does not always have that the 2-form $d\alpha$ will also annihilate any such tangent vector, if one is to define the exterior derivative on transverse $k$-forms in general, it would be more useful to also require that $d\alpha$ will also annihilate any such tangent vector. Hence, we define a 1-form $\alpha$ to be *basic* iff $i_\mathbf{u}\alpha = 0$ and $i_\mathbf{u}d\alpha = 0$ for any $\mathbf{u} \in \mathfrak{X}(\mathcal{O})$. Note that this is stronger than requiring that $L_\mathbf{u}\alpha = 0$ for any $\mathbf{u} \in \mathfrak{X}(\mathcal{O})$ since:

$$L_\mathbf{u}\alpha = di_\mathbf{u}\alpha + i_\mathbf{u}d\alpha = 0 \qquad (4.4)$$

for any basic 1-form $\alpha$, but the vanishing of $L_\mathbf{u}\alpha$ does not have to imply that both of the terms in the sum vanish identically.

Once again, the condition that $\alpha$ be basic is equivalent to the condition that its component functions in any adapted local coordinate system must be functions only of the spatial coordinates. We also have that the set $\Lambda_b^1(M)$ of all basic 1-forms on $M$ is a $C_b^\infty(M)$–module.

Since the Lie derivative acts on tensor products of vector fields and covector fields as a derivation, one easily sees that tensor products of basic objects are again basic in the sense of being constant along the leaves of $\mathcal{O}$ with respect to Lie derivation. Hence, we add the subscript $b$ to the usual notations to refer to the spaces of basic tensor fields on $M$ of various types.

In particular, the exterior product of $k$ basic 1-forms is a basic $k$-form in the same sense as defined for 1-forms. We can then define the exterior algebra $\Lambda_b^*(M) = \bigoplus_{k=0}^{3} \Lambda_b^k(M)$ of basic forms on $M$. The exterior algebra of basic forms is then a $C_b^\infty(M)$–module. Note that since we are essentially generalizing the pull-backs of $k$-forms on $\Sigma$ by $p$, as above, and $\Sigma$ is three-dimensional, the basic 4-forms must vanish.

By definition, the exterior derivative of a basic $k$-form will be a basic $k+1$-form. Hence, one can define closed and exact basic forms in the predictable way, as well as the basic de Rham cohomology $H_b^*(M; \mathbb{R})$. However, one must then realize that the basic cohomology is more pertinent to the topology of the leaf space than the topology of $M$.

If the leaf space $M/\mathcal{O}$ were a differentiable manifold $\Sigma$ then it would have a tangent bundle $T(\Sigma)$ and one could pull it back by means of the projection $p:M \to \Sigma$ to a vector bundle on $M$. However, such a pull-back would not actually be a sub-bundle of $T(M)$ since any tangent vector to $M$ that is transverse of $\mathcal{O}$ will project to a non-zero vector tangent to Σ. That is, one associates a vector tangent to $p(x)$ with $x$, not a vector tangent to $x$. In the more general case, for which $p$ is not necessarily a submersion, the vector bundle that one works with as a substitute for the "spatial tangent bundle" is the *transverse vector bundle* $Q(M)$ to $\mathcal{O}$.

A *transverse (or normal) vector* $\overline{\mathbf{X}}$ to $\mathcal{O}$ at $x \in M$ is an equivalence class of all tangent vectors $\mathbf{v}, \mathbf{w} \in T_x(M)$ such that $\mathbf{v} - \mathbf{w} \in T_x(\mathcal{O})$. One can form linear combinations of such equivalence classes in the obvious way, so the set $Q_x(M)$ of all transverse vectors to $\mathcal{O}$ at $x \in M$ can be given the structure of a vector space. Similarly, the disjoint union $Q(M)$ of all these transverse vector spaces for all $x$ can be given the structure of a vector bundle over $M$ whose rank is complementary to that of $T(\mathcal{O})$, namely, three in the present case. However, once again, although the equivalence classes consist of tangent vectors, $Q(M)$ is not itself a sub-bundle of $T(M)$. The canonical projection $q: T(M) \to Q(M)$, $\mathbf{X} \mapsto \overline{\mathbf{X}}$ behaves in a manner that is analogous to that of the projection of $T(M)$ onto $T(\Sigma)$ in the case of a simple foliation. Hence, the transverse bundle $Q(M)$ will be the starting point for the rest of transverse geometry.

We can define a *transverse vector field* to be a section $\overline{\mathbf{X}}: M \to Q(M)$ of the fibration $Q(M) \to M$, and denote the set of all transverse vector fields by $\mathfrak{X}(Q)$. Note that although the dimension of the fibers of $Q(M)$ is one less than the dimension of $M$, one could not think of a transverse vector field as "spatial" vector field unless it were constant along the orbits; i.e., unless it were a foliate vector field. One thus has a projection $\mathfrak{X}_\mathcal{O}(M) \to \mathfrak{X}(Q)$ whose kernel is $\mathfrak{X}(\mathcal{O})$. One can then think of a transversal vector field as representing essentially a "time-varying spatial vector field" and a foliate vector field as "time-invariant," and therefore essentially spatial with respect to the observer $\mathcal{O}$ (Compare this with the methodology of "parametric manifolds" that is described by Perjés [**19**] and in Boersma and Dray [**5**].)

The set $\mathfrak{X}(Q)$ can be given not only the structure of an infinite-dimensional real vector space in the obvious way, but also the structure of an infinite-dimensional real Lie algebra by the definition:

$$[\overline{\mathbf{X}}, \overline{\mathbf{Y}}] = q[\mathbf{X}, \mathbf{Y}], \tag{4.4}$$

in which $\mathbf{X}, \mathbf{Y} \in \mathfrak{X}_\mathcal{O}(M)$ are any foliate vector fields that project onto $\overline{\mathbf{X}}, \overline{\mathbf{Y}}$, resp.

It is immediate that if one defines the action of a transverse vector field $\overline{\mathbf{X}}$ on a basic function $f$ to be:

$$\overline{\mathbf{X}} f = \mathbf{X} f \tag{4.5}$$

for any vector field $\mathbf{X} \in \mathfrak{X}(M)$ that projects to $\overline{\mathbf{X}}$ then this definition is unambiguous.

Note that we can now characterize a foliate vector field $\mathbf{X}$ on $M$ by the property that:

$$q[\mathbf{X}, \mathbf{u}] = 0 \tag{4.6}$$

for any $\mathbf{u} \in \mathfrak{X}(\mathcal{O})$. One can think of the expression on the left-hand side as essentially the transverse Lie derivative of $\mathbf{X}$ along the foliation $\mathcal{O}$, since any other choice of $\mathbf{u}$ will give the same result.

It is important to point out that the evaluation of a basic $k$-form on a set of transverse vector fields does not have to produce a basic function. Hence, one loses the interpretation of $k$-forms as $C_b^\infty(M)$-multilinear functionals on transverse multivector fields. In particular, the space $\Lambda_b^1(M)$ of basic 1-forms is not generally the dual vector space to $\mathfrak{X}(Q)$.

Since the fibers of $Q(M)$ are three-dimensional vector spaces they can be framed by linear 3-frames. A *transverse* 3-frame $\{\bar{\mathbf{e}}_i, i = 1, 2, 3\}$ at $x \in M$ then consists of three linearly independent transverse vectors $\bar{\mathbf{e}}_i$ in $Q_x(M)$. Equivalently, one can regard a transverse frame as an isomorphism $\sigma: \mathbb{R}^3 \to Q_x(M)$, $v^i \mapsto v^i \bar{\mathbf{e}}_i$. One can also think of a transverse 3-frame at $x$ as an equivalence class of 3-frames $\mathbf{e}_i$ in $T_x(M)$ that are transversal to $\mathcal{O}$, in the sense that the subspace of $T_x(M)$ that they span is complementary to $T_x(\mathcal{O})$, and project onto the corresponding transverse vectors $\bar{\mathbf{e}}_i$ under $q$. There is also a well-defined projection from adapted linear 4-frames in $T_x(M)$ to transverse 3-frames since the frame member that is tangent to $\mathcal{O}$ will vanish under the projection $q$.

One can then define the $GL(3)$-principal bundle $GL(Q)$ of all transverse frames, and one then calls it the *transverse frame bundle* for $\mathcal{O}$.

Once again, if we have a local transverse 3-frame field $\bar{\mathbf{e}}_i: U \to GL(Q)$ then we must note that there is a difference between representing it by means of three vector fields $\mathbf{e}_i: U \to T(M)$ that project onto $\bar{\mathbf{e}}_i$, which is the time-varying case, and three foliate vector fields that project in that manner, which is then the time-invariant case. It is only in the latter case that one is dealing with a "projectable" transverse 3-frame field. Since our foliation is one-dimensional, we can then characterize such 3-frame fields by saying that for any temporal vector field $\mathbf{u}$ one must have that the three vector fields $L_\mathbf{u} \bar{\mathbf{e}}_i$ are all temporal. One can then regard the triple of foliate vector fields on $U$ as representing a spatial 3-frame relative to $\mathcal{O}$.

Such a projectable transverse 3-frame can be obtained locally when $\mathcal{O}$ is time-orientable and defined by a non-zero vector field $\mathbf{u}$. If $\Phi: (-\varepsilon, +\varepsilon) \times U \to M$ is a local flow for $\mathbf{u}$ on $U \subset M$ then if $\Phi$ has a slice $S \subset U$, which we assume to represent $\tau = 0$, and a 3-frame field $\bar{\mathbf{e}}_i$ on $S$ then one can extend $\bar{\mathbf{e}}_i$ to the rest of the image of $\Phi$ by convection. That is: one pushes the frame $\bar{\mathbf{e}}_i$ forward along the flow of $\mathbf{u}$ by means of the diffeomorphisms of the flow itself. Such a construction is always possible when $U$ is the domain of an adapted coordinate chart since one can then use any level set of the $x^0$ coordinate as a slice. We shall then call a local transverse frame field that is obtained in this manner a *convected local transverse frame field*.

Now that we have defined a principal bundle of frames over $M$, we can then go about the usual business of differential geometry, as it is practiced on frame bundles (see

Kobayashi and Nomizu [**16**], Bishop and Crittenden [**3**], or Sternberg [**21**]), by starting with $GL(Q)$ as if it were the bundle $GL(\Sigma)$ of linear 3-frames on a spatial manifold $\Sigma$.

First, one notes that $GL(Q)$ has a canonical 1-form $\bar{\theta}^i$ with values in $\mathbb{R}^3$ that works essentially the same way as the canonical $\mathbb{R}^4$-valued 1-form $\theta^\mu$ on $GL(M)$. That is, if $x \in M$ and $\bar{\mathbf{e}}_i \in GL_x(Q)$ is a transverse 3-frame at $x$ then $\bar{\mathbf{e}}_i$ has a reciprocal coframe $\bar{\theta}^i$, which, by definition, then has the property that $\bar{\theta}^i(\bar{\mathbf{e}}_j) = \delta^i_j$. One can also think of $\bar{\theta}^i$ as a linear isomorphism $\bar{\theta}^i : Q_x(M) \to \mathbb{R}^3$, which is the inverse of the isomorphism defined by $\bar{\mathbf{e}}_i$, or, for that matter, an $\mathbb{R}^3$-valued 1-form on $Q_x(M)$. One then pulls this 1-form up to an $\mathbb{R}^3$-valued 1-form on $T_{\bar{\mathbf{e}}}GL_x(Q)$ that we also denote by $\bar{\theta}^i$. It can also be characterized by:

$$\bar{\theta}^i_{\bar{\mathbf{e}}}(\mathbf{X}) = \bar{\theta}^i_x(\overline{p_*(\mathbf{X})}), \qquad (4.7)$$

in which the overbar on the right-hand side refers to the projection of the vector $p_*(\mathbf{X})$ $T_x(M)$ into $Q_x(M)$.

By construction, the resulting 1-form $\bar{\theta}^i$ on $GL(Q)$ will be $GL(3)$-invariant and if $\bar{\mathbf{e}}_i : U \to GL(Q)$ is a local transverse 3-frame field then the pull-down of $\bar{\theta}^i$ to $U$ by way of $\bar{\mathbf{e}}_i$ will be the reciprocal coframe field to $\bar{\mathbf{e}}_i$.

A *transverse linear connection* is a linear connection on the bundle $GL(Q)$. Hence, it can be defined by a $GL(3)$-invariant horizontal sub-bundle $H(Q)$ in $T(GL(Q))$ that is complementary to the vertical sub-bundle $V(Q)$, which consists of all tangent vectors to $GL(Q)$ that project to zero under the differential of the bundle map $GL(Q) \to M$. A transverse linear connection can also be defined by an $\text{Ad}^{-1}$-equivariant vertical 1-form $\bar{\omega}^i_j$ on $GL(Q)$ with values in $\mathfrak{gl}(3)$. The horizontal subspaces on $GL(Q)$ then become the annihilating subspaces of this 1-form.

If $\mathcal{O}$ were a simple foliation, so the projection $p:M \to M/\mathcal{O}$ would be a submersion of manifolds, then the bundle $GL(Q)$ would be the pull-back of the bundle $GL(M/\mathcal{O})$ of linear 3-frames on the manifold $M/\mathcal{O}$ and a transverse linear connection $\bar{\omega}^i_j$ would be the pull-back of some corresponding linear connection on $GL(M/\mathcal{O})$ by $p$, just as the fundamental 1-form $\bar{\theta}^i$ would be the pull-back of the corresponding fundamental 1-form on $GL(M/\mathcal{O})$. Hence, we shall call a transverse linear connection on $GL(Q)$ *projectable* iff it has that property on any simple open subset of $M$; i.e., every open subset over which the restriction of the foliation is simple. In general, transverse linear connections do not have to be projectable, but, as it turns out, the ones that we shall be concerned with shortly – viz., *transverse Riemannian connections* – are always projectable.

Given a local transverse frame field $\bar{\mathbf{e}}_i : U \to GL(Q)$ one can pull down $\bar{\omega}^i_j$ to a basic 1-form on $U$ with values in $\mathfrak{gl}(3)$. If $\bar{\mathbf{e}}_i$ represents the natural frame field for an adapted local coordinate system then one can express $\bar{\omega}^i_j$ in terms of the coordinate 1-forms as:

$$\bar{\omega}^i_j = \bar{\Gamma}^i_{jk} \bar{\theta}^k . \tag{4.8}$$

for appropriate basic functions $\bar{\Gamma}^i_{jk}$.

Although the *Cartan structure equations* for the connection $\bar{\omega}^i_j$ can be deduced as consequences of other definitions for the *torsion 2-form* $\bar{\Theta}^i$ and *curvature 2-form* $\bar{\Omega}^i_j$ on *GL*(*Q*), we shall simply use those equations as the definitions themselves.

$$\bar{\Theta}^i = \bar{\nabla}\bar{\theta}^i = d\bar{\theta}^i + \bar{\omega}^i_j \wedge \bar{\theta}^j , \qquad \bar{\Omega}^i_j = \bar{\nabla}\bar{\omega}^i_j = d\bar{\omega}^i_j + \bar{\omega}^i_k \wedge \bar{\omega}^k_j . \tag{4.9}$$

The 2-forms thus defined satisfy the usual *Bianchi identities:*

$$\bar{\nabla}\bar{\Theta}^i = d\bar{\Theta}^i + \bar{\omega}^i_j \wedge \bar{\Theta}^j = \bar{\Omega}^i_j \wedge \bar{\theta}^j , \qquad \bar{\nabla}\bar{\Omega}^i_j = d\bar{\Omega}^i_j + \bar{\omega}^i_k \wedge \bar{\Omega}^k_j = 0. \tag{4.10}$$

A connection on *GL*(*Q*) allows one to define a notion of the parallel translation of a transverse 3-frame $\bar{\mathbf{e}}_i$ along a curve $\gamma(s)$. If $\mathbf{v}(s)$ is the velocity vector field of $\gamma$ and $\bar{\mathbf{e}}_i(s)$ is a transverse 3-frame field along $\gamma$ then one can think of $\bar{\mathbf{e}}_i(s)$ as a differentiable curve in *GL*(*Q*) that projects onto $\gamma$. If the velocity vector field $\dot{\bar{\mathbf{e}}}_i(s)$, which is called the *lift* of $\mathbf{v}(s)$, is horizontal, so:

$$\bar{\omega}^i_j(\dot{\bar{\mathbf{e}}}_i(s)) = 0 \tag{4.11}$$

then one can regard the transverse frame field $\bar{\mathbf{e}}_i(s)$ as *parallel along* $\gamma$.

One can also define a transverse linear connection by means of a *covariant differential* operator on the transverse vector fields:

$$\bar{\nabla} : \mathfrak{X}(Q) \to T^*(M) \otimes \mathfrak{X}(Q), \quad \bar{\mathbf{Y}} \mapsto \bar{\nabla}\bar{\mathbf{Y}} . \tag{4.12}$$

The reason for the appearance of $T^*(M)$ is the fact that the ordinary differential map for a section $\bar{\mathbf{Y}} : M \to Q(M)$, takes $T(M)$ to $T(Q)$.

The easiest way to relate such a definition to the previous one is to choose a local transverse 3-frame field $\bar{\mathbf{e}}_i$, so $\bar{\mathbf{Y}} = \bar{Y}^i \bar{\mathbf{e}}_i$ and:

$$\bar{\nabla}\bar{\mathbf{Y}} = (d\bar{Y}^i + \bar{\omega}^i_j \bar{Y}^j)\bar{\mathbf{e}}_i . \tag{4.13}$$

The evaluation of $\bar{\nabla}\bar{\mathbf{Y}}$ on a vector field $\mathbf{X}$ then produces a transverse vector field $\bar{\nabla}_{\mathbf{X}}\bar{\mathbf{Y}}$ that represents the *covariant derivative* of $\bar{\mathbf{Y}}$ in the direction of $\mathbf{X}$. Its local form is:

$$\bar{\nabla}_{\mathbf{X}}\bar{\mathbf{Y}} = (\mathbf{X}\bar{Y}^i + \bar{\Gamma}^i_{jk} X^j \bar{Y}^k)\bar{\mathbf{e}}_i . \tag{4.14}$$

We can then characterize a transverse frame field $\bar{\mathbf{e}}_i(s)$ along a smooth curve $\gamma$ as being parallel along that curve iff:

$$\bar{\nabla}_{\mathbf{v}(s)} \bar{\mathbf{e}}_i(s) = 0 \qquad i = 1, 2, 3, \text{all } s \tag{4.15}$$

which has the local form:

$$i_{\mathbf{v}} D\bar{\mathbf{e}}_i = \bar{\omega}_i^j(\mathbf{v}) \bar{\mathbf{e}}_j. \tag{4.16}$$

If we wish to define transverse geodesics for this connection, we must be careful to note the subtlety associated with the difference between the velocity vector field $\mathbf{w}(s)$ to a curve $\chi(s)$ in $M$ and its projection onto a transverse vector field $\bar{\mathbf{w}}(s)$. Many tangent vectors in each $T_{\chi(s)}(M)$ will project onto the same transverse vector, so although one can define the differential equation to be satisfied along such a curve in the predictable way, if one defines the differential equation for a geodesic vector field in the same manner then its integrability becomes ambiguous. Hence, we shall define transverse geodesics only for the Riemannian case later on.

A transverse covariant derivative that figures crucially in the study of characteristic classes of foliated bundle is the *Bott connection*. It is defined by means of:

$$\overset{o}{\nabla}_{\mathbf{X}} \bar{\mathbf{Y}} = p[\mathbf{X}, \mathbf{Y}] = p \mathrm{L}_{\mathbf{X}} \mathbf{Y}, \tag{4.17}$$

in which $\mathbf{X} \in \mathfrak{X}(M)$ is arbitrary, $\bar{\mathbf{Y}} \in \mathfrak{X}(Q)$, and $\mathbf{Y} \in \mathfrak{X}(M)$ is any vector field that projects onto $\bar{\mathbf{Y}}$. This definition is unambiguous because the effect of choosing a different $\mathbf{Y}$ will only add a temporal vector field, which then projects to zero. One can then think of the Bott connection as defining essentially a transverse Lie derivative operator on transverse vector fields. As a consequence of the Jacobi identities, and the definition of curvature in terms of covariant derivatives:

$$\overset{o}{R}(\mathbf{X}, \mathbf{Y})\mathbf{Z} = \overset{o}{\nabla}_{\mathbf{X}} \overset{o}{\nabla}_{\mathbf{Y}} \mathbf{Z} - \overset{o}{\nabla}_{\mathbf{Y}} \overset{o}{\nabla}_{\mathbf{X}} \mathbf{Z} - \overset{o}{\nabla}_{[\mathbf{X},\mathbf{Y}]} \mathbf{Z} \tag{4.18}$$

it will have vanishing curvature.

If $G$ is a subgroup of $GL(3)$ then we can consider reducing the bundle $GL(Q)$ to a bundle $G(Q)$ of transverse 3-frames that all lie with the orbits of the action of $G$ on $GL(Q)$. Such a reduction of $GL(Q)$ to a $G$-principle bundle is called a *transverse $G$-structure*. For the sake of space-time structure, the most elementary subgroups to consider are $GL^+(3)$, $SL(3)$, $SO(3)$, and $\{e\}$, the identity subgroup.

The subgroup $GL^+(3)$ consists of invertible 3×3 real matrices with positive determinant, which then have the property that they preserve a choice of orientation on $\mathbb{R}^3$. A transverse $GL^+(3)$-structure on $M$ then represents a *transverse orientation*, if it exists; i.e., an orientation on the bundle $Q(M)$. If $T(M)$ and $T(\mathcal{O})$ are both orientable then it follows that $Q(M)$ is orientable.

The subgroup *SL*(3) consists of invertible 3×3 real matrices with unit determinant, which then have the property that they preserve the unit volume element $\mathbf{e}_1 \wedge \mathbf{e}_2 \wedge \mathbf{e}_3$ on $\mathbb{R}^3$. Hence, a transverse *SL*(3)-structure on *M* then represents a choice of unit-volume element on *Q*(*M*), which then can be regarded as a *transverse unit volume element*. If *Q*(*M*) is orientable then the reduction to a transverse *SL*(3)-structure is automatic, since *SL*(3) is a deformation retract of $GL^+(3)$.

Of course, *SO*(3) represents the subgroup of *SL*(3) that consists of (Euclidian) orthogonal matrices. Hence, a transverse *SO*(3)-structure *O*(*Q*) on *M* is equivalent to a choice of Riemannian metric $\bar{g}$ on *Q*(*M*), for which *O*(*Q*) becomes the bundle of orthonormal frames. In this case, $\mathcal{O}$ is commonly called a *Riemannian foliation.*

Finally, a transverse {*e*}-structure on *M* is a choice of a unique transverse 3-frame at each of its points, which one then calls a *transverse parallelism.* One also calls says that such a foliation is *framed.* Although one usually thinks of the global parallelizability of a manifold as something that is hard to come by, even for homogeneous spaces, and especially for dimension four, it is intriguing that in the case of dimension three one has the useful fact that any compact orientable 3-manifold is parallelizable. Hence, if one is attempting to generalize compact orientable spatial manifolds then the possibility of *Q*(*M*) being trivializable is worth considering.

Let us return to the case of Riemannian foliations. One notes that if the foliation $\mathcal{O}$ were simple then the pull-back $\bar{g} = p^*g$ of a Riemannian metric *g* on $M/\mathcal{O}$ would have to be degenerate because vectors tangent to the leaves of $\mathcal{O}$ would project to zero. In relativity, one naturally wishes to know how one could obtain $\bar{g}$ by the projection of a non-degenerate metric, such a Lorentzian metric. In the terminology of Reinhart [**20, 21**], a metric *g* on *T*(*M*) is called *bundle-like* iff whenever $\mathbf{X}, \mathbf{Y} \in \mathfrak{X}_\mathcal{O}(U)$ are foliate vector fields on any open set *U* that are orthogonal to the leaves of $\mathcal{O}$ the function *g*(**X**, **Y**) on *U* is basic. Hence, such a metric on *T*(*M*) is essentially constant on the leaves of $\mathcal{O}$ when restricted to the tangent spaces that are orthogonal to the leaves.

One has the usual consequences of defining a transverse metric $\bar{g}$ on *Q*(*M*): There is a unique metric connection with vanishing torsion associated with $\bar{g}$ that one calls the *transverse Levi-Civita connection.* For a simple foliation, it is the pull-back of the Levi-Civita connection on $O(M/\mathcal{O})$ that is defined by the projected metric $\bar{g}$ on $T(M/\mathcal{O})$. Hence, the transverse Levi-Civita connection is a projectable connection on *O*(*Q*).

We define transverse geodesics for this connection by means of a bundle-like metric *g* on *T*(*M*) that locally projects to $\bar{g}$. That is, a *transverse geodesic* for the transverse Levi-Civita connection associated with $\bar{g}$ is a geodesic for the Levi-Civita connection of *g* that is orthogonal to the leaves of $\mathcal{O}$. Such a transverse geodesic has the following two properties, which also completely characterize such curves:

*i.* If a geodesic of *g* is orthogonal to a leaf of $\mathcal{O}$ at one point then it is it orthogonal to the leaves at all of its other points; i.e., it is a transverse geodesic.

*ii*. If *U* is a simple open subset then the intersection of a transverse geodesic with *U* projects onto a geodesic in $U/\mathcal{O}$ for the projected metric.

These properties were also established by Reinhart [**20**].

For the sake of cosmological models – certainly, the case of Bianchi cosmologies – which invariably involve the existence of Killing vector fields and the action of isometry groups, it is important to define what such objects turn into when one no longer assumes that there is a spatial manifold $\Sigma$ and a product foliation $\mathbb{R} \times \Sigma$ of space-time. The definition is straightforward: A vector field **X** on *M* is a *transverse Killing vector field* for the transverse metric $\bar{g}$ iff:

$$L_{\mathbf{X}} \bar{g} = 0. \tag{4.19}$$

In this equation, we clarify that $\bar{g}$ is also represented by the degenerate metric on $T(M)$:

$$\bar{g}(\mathbf{X}, \mathbf{Y}) \equiv \bar{g}(\bar{\mathbf{X}}, \bar{\mathbf{Y}}). \tag{4.20}$$

The flow of a transverse Killing vector field then consists of isometries for the metric $\bar{g}$.

A basic property of transverse Killing vector fields is that they are always foliate. (See, Molino [**18**].) Note that this does not imply that they must always be transverse to the given one-dimensional foliation $\mathcal{O}$, only that they be constant along it in the Lie sense. Whenever a transverse Killing vector field is tangent to $\mathcal{O}$ the tangent vector at that point will then project to zero in $Q(M)$, which defines a fixed point of the local flow of isometries.

Since most cosmological models – in particular, Bianchi cosmologies [**8, 28**] – involve the imposition of the action of a group of isometries on the spacetime, we see that the isometry groups that can be represented by transverse Killing vector fields are primarily relevant to the assumptions concerning the spatial homogeneity, isotropy, and self-similarity of the model in question. However, if the orbit space (i.e., the surfaces of homogeneity) of a transverse group action defines a Lie foliation, in which case all of the isotropy subgroups must have the same dimension, one cannot have orbits that are both three-dimensional and transverse to $\mathcal{O}$ unless the foliation $\mathcal{O}$ is spatially integrable (i.e., hypersurface normal).

## 5. Relationship to 1+3 splittings.

A more established manner of obtaining spatial geometry from space-time geometry when one is given a congruence of curves defined by a physical observer $\mathcal{O}$ is to complete the line bundle $L(M)$ that is tangent to the congruence with a rank three sub-bundle $\Sigma(M)$ so that one obtains a Whitney sum splitting of $T(M)$ into $L(M) \oplus \Sigma(M)$. Such a decomposition is then called a *1+3 splitting* of $T(M)$, or a "threading" of space-time [**4, 9, 14, 15**], to distinguish it from the complementary case, in which one has a "slicing" of space-time by a codimension-one foliation of proper-time simultaneity hypersurfaces; in such a case, one refers to a *3+1 splitting* [**4**, **6**, **15**] Once again, the advantage of a threading over a slicing is most meaningful when one is dealing with a spatially non-integrable congruence; i.e., one that is not hypersurface normal.

Because the fibers of $\Sigma(M)$, in either event, will be transverse to $L(M)$ – i.e., transverse to the foliation – one sees that the restriction of the projection $p: T(M) \to Q(M)$ to $\Sigma(M)$ will be a vector bundle isomorphism. Because of this isomorphism, one can think of any choice of $\Sigma(M)$ as basically a "faithful representation" of the transverse bundle $Q(M)$. Hence, a 1+3 splitting of $T(M)$ plays essentially the same role in regard to $Q(M)$ that a choice of frame plays in regard to a fiber of $T(M)$.

When $M$ is given a Lorentzian structure in addition to $\mathcal{O}$, the natural choice of 1+3 splitting is an orthogonal splitting, for which the fibers of $\Sigma(M)$ are the orthogonal complements to the lines of $L(M)$. In the event that the sub-bundle $\Sigma(M)$ is non-integrable, it is important to point out that one cannot have an adapted local coordinate chart $(U, x^\mu)$ whose natural frame field $\partial_\mu$ is also orthonormal for the metric. That is because the tangent spaces to the local foliation of $U$ defined by the level surfaces of $x^0$ would have to agree with the fibers of $\Sigma(U)$, but the local foliation would necessarily be integrable. Hence, one must represent the fibers of $\Sigma(U)$ as the annihilating hyperspaces of a timelike 1-form $M$ on $U$:

$$M = M_0 \, dx^0 + M_i \, dx^i \tag{5.1}$$

that does not coincide with $dx^0$. Up to sign and normalization, the $M_0$ component of this 1-form is called the *lapse* function and the components $M_i$ define the *shift* covector for the threading defined by $\mathcal{O}$.

We point out that the methodology of transverse geometry also has an immediate relevance to the geometry of "anholonomic spaces" (see, Vranceanu [**27**], Horak [**13**], Synge [**25**]), or Schouten [**22**]), which also start with 1+3 splittings of $T(M)$ for which the $\Sigma(M)$ sub-bundle is non-integrable, as well as "nonlinear connections" (see, Vacaru, et al. [**26**], Bejancu and Farran [**2**], or Delphenich [**7**] and references therein). In the methods of nonlinear connections, one regards a 1+3 splitting of $T(M)$ as a generalization of the splitting of the tangent bundle to a principal fiber bundle into a horizontal and vertical sub-bundle that one uses to define a connection.

## 6. Discussion.

We have seen that when one is dealing with a spatially non-integrable cosmological model $(M, g, \mathbf{u})$, viz., one for which the vorticity vector field of $\mathbf{u}$ is non-vanishing, it is still possible to speak of the spatial geometry of $M$ relative to the physical observer $\mathbf{u}$ by employing the methodology of the transverse geometry of foliations, suitably specialized to the case of one-dimensional foliations. It is not necessary to either introduce a complementary sub-bundle to the tangent line bundle generated by $\mathbf{u}$ or a transverse foliation. Indeed, although such a complementary sub-bundle will always exist, the existence of a transverse foliation would imply the spatial integrability of $\mathbf{u}$.

Some obvious directions to pursue in the name of adapting transverse geometry to the established methods of relativity and cosmology would be:

*i*. Detailing the form that decomposition of the Riemannian curvature tensor into the Weyl curvature tensor, the Ricci part, and the scalar curvature would take, especially the form of the electric and magnetic parts of the Weyl tensor, since they are commonly introduced to account for spatial tidal gravitational forces.

*ii*. Investigating the corresponding form that geodesic deviation takes, including the aforementioned decomposition of curvature.

*iii*. Formulating the Newtonian limit in terms of transverse geometry.

*iv*. Examining the form that the Bianchi classification scheme for homogeneous cosmologies takes in transverse geometry.

*v*. Applying the methodology to some existing solution with non-vanishing vorticity, such as the Gödel solution.

*vi*. Seeing whether the formalism leads to any new spatially non-integrable solutions.

Nevertheless, in advance of such extensions, one can still see that the formalism of transverse geometry seems to be a natural addition to the differential geometry of spacetime.